\documentclass[preprint]{elsarticle}

\usepackage{graphicx}
\usepackage{amssymb}
\usepackage{tabularx}
\usepackage{xcolor}
\journal{Journal of Magnetism and Magnetic Materials}

\begin{document}

\begin{frontmatter}

\title{The Co$^{2+}$-Ir$^{5+}$ orbital hybridization in LaCaCoIrO$_6$ double perovskite}

\author[AA]{L. Bufai\c{c}al\corref{Bufaical}}, \ead{lbufaical@ufg.br}, \author[BB]{L. S. I. Veiga},\author[CC]{J. R. L. Mardegan}, \author[CC]{T. Pohlmann}, \author[CC]{S. Francoual}, \author[BB]{S. S. Dhesi}, \author[DD]{M. B. Fontes}, \author[DD]{E. M. Bittar}

\address[AA]{Instituto de F\'{i}sica, Universidade Federal de Goi\'{a}s, 74001-970 , Goi\^{a}nia, GO, Brazil}

\address[BB]{Diamond Light Source, Chilton, Didcot, Oxfordshire, OX11 0DE, United Kingdom}

\address[CC]{Deutsches Elektronen-Synchrotron DESY, Notkestra$\beta$e 85, 22607, Hamburg, Germany}

\address[DD]{Centro Brasileiro de Pesquisas F\'{\i}sicas, Rua Dr. Xavier Sigaud 150, 22290-180, Rio de Janeiro, RJ, Brazil}

\cortext[Bufaical]{Corresponding author}

\begin{abstract}

Here, we report on the structural, electronic, and magnetic properties of a polycrystalline sample of the LaCaCoIrO$_6$ double-perovskite investigated by means of synchrotron x-ray powder diffraction, x-ray absorption spectroscopy, and x-ray magnetic circular dichroism at the Co and Ir $L_{2,3}$ edges, magnetometry, and electrical transport. Our results indicate a configuration of nearly Co$^{2+}$/Ir$^{5+}$ configuration for the transition-metal ions, with spin canting within the Co antiferromagnetic superstructure responsible for the ferromagnetic-like behavior observed below 100 K. The highly insulating character of LaCaCoIrO$_6$ and its positive magnetoresistance further suggest that this antiferromagnetic superexchange interaction occurs through an indirect hybridization between the Co $e_g$ orbitals.

\end{abstract}


\end{frontmatter}

\section{Introduction}

There has recently been a great attention directed towards iridates driven by the the peculiar interplay between the spin-orbit coupling (SOC), the crystal field and the on-site Coulomb repulsion. This interplay leads to the splitting of the Ir $t_{2g}$ levels, resulting in a lower-energy fourfold state with $j_{eff}$ = 3/2 and an upper-energy twofold state with $j_{eff}$ = 1/2 \cite{Rotenberg}. In Ir$^{4+}$ (5$d^{5}$)-based oxides it has been reported, for instance, a $j_{eff}$ = 1/2 Mott insulating state in Sr$_2$IrO$_4$ \cite{Kim}, and a possibly Kitaev spin liquid state in Cu$_2$IrO$_3$ \cite{Abramchuk}. But unusual magnetism can also occur in Ir$^{5+}$-based systems. For example, there is an ongoing discussion concerning a possible excitonic magnetic ground state for Ir 5$d^{4}$ in Sr$_2$YIrO$_6$ and Ba$_2$YIrO$_6$ double perovskites (DP), which from a $j_{eff}$ perspective should result in a $J$ = 0 state \cite{Cao,Cao2,Irifune}.

The addition of a second transition-metal (TM) ion in Ir-DPs increases the complexity of the system's electronic and magnetic structures. In contrast to 3$d$-3$d$ systems where the main exchange interactions involve the $e_g$ orbitals, for 3$d$-5$d$ systems may involve the $t_{2g}$ orbitals. As a result, the conventional Goodenough-Kanamori-Anderson rules often fail to describe the magnetic coupling between the TM ions \cite{Kanungo,Morrow,Wu}. A recent study on LaSrNiIrO$_6$, for instance, has shown an antiferromagnetic (AFM) superstructure at the Ni$^{2+}$ site through the Ni$^{2+}$--O--Ir$^{5+}$--O--Ni$^{2+}$ path \cite{Hayward}. On the other hand, the complete substitution of Sr$^{2+}$ by a rare-earth A$^{3+}$ element in A$_2$NiIrO$_6$ DPs leads to a magnetic Ir$^{4+}$ state, which couples AFM with the Ni$^{2+}$ ions to produce a ferrimagnetic (FIM) behavior whose strength increases as the A-site ionic radius decreases \cite{Ferreira}. Because the Ni$^{2+}$ (3$d^{8}$) $t_{2g}$ orbitals are completely filled, the hybridization with Ir $j_{eff}$ = 1/2 necessarily occurs via its half-filled $e_g$ orbitals \cite{Feng}. 

Another example of interesting Ir$^{5+}$-based DP is Sr$_2$FeIrO$_6$, for which an AFM coupling between the Fe$^{3+}$ ions occurs at $T_N$ $\simeq$ 120 K \cite{Battle,JSSC,Page}. This suggests strong magnetic exchange between the widely separated Fe $d$ ions through the Fe--O--Ir--O--Fe pathway. Here, Fe$^{3+}$ presents empty states at both $t_{2g}$ and $e_g$ orbitals, and based on existing repors, it remains unclear whether the $t_{2g}$ orbitals or the $e_g$ orbitals are more favorable for hybridization with Ir$^{5+}$. The same situation holds for LaSrCoIrO$_6$, for which neutron powder diffraction (NPD) and x-ray magnetic circular dichroism (XMCD) confirm an AFM arrangement for the Co ions, but the preferable path of hybridization between the TM ions remains unknown \cite{Narayanan,Kolchinskaya}. 

Thereby, compelling question for this work is what constitutes the most relevant channel of hybridization between the TM ions in a scenario where Ir$^{5+}$ is the 5$d$ element and the 3$d$ ion presents empty states in both $t_{2g}$ and $e_g$ orbitals. Our case study is the LaCaCoIrO$_6$ (LCCIO) DP, for which the presumably Co$^{2+}$/Ir$^{5+}$ configuration leads to two fully occupied $t_{2g}$ orbitals and one half-filled $t_{2g}$ orbital for the high spin (HS) Co$^{2+}$ (3$d^{7}$: $t_{2g}^5e_g^2$) ion. Conversely, for the low spin (LS) Ir$^{5+}$ (5$d^{4}$: $t_{2g}^4e_g^0$) the four valence electrons occupy lower $j_{eff}$ = 3/2, leaving the $j_{eff}$ = 1/2 levels unoccupied. We thoroughly investigate the structural, electronic, and magnetic properties of a polycrystalline sample of LCCIO by means of synchrotron x-ray powder diffraction (SXRD), x-ray absorption spectroscopy (XAS), and XMCD at the Co and Ir $L_{2,3}$-edges, magnetometry, and electrical transport measurements to unravel the main exchange interactions between the TM ions present within this system.

\section{Experimental details}

The polycrystalline LCCIO sample was synthesized by conventional solid-state reaction, as described elsewhere \cite{PRB2020}. High-resolution SXRD data were recorded on beamline XPD at the Brazilian Synchrotron Light Laboratory (LNLS). The SXRD patterns were obtained using the Bragg-Brentano geometry with wavelength $\lambda$ = 1.3736 $\textrm{\AA}$, at temperatures ranging from 300 K down to 25 K, using a DE-202 cryostat (ARS Cryo) with a HOPG(002) analyzer. Rietveld refinements were performed using the GSAS software and its graphical interface program \cite{GSAS}. 

Room temperature Co $L_{2,3}$-edge XAS was carried on beamline I06 at Diamond Light Source (DLS), where a fine powder of LCCIO was spaded on conductive carbon tape and measured in total electron yield (TEY). The XMCD measurements at the same absorption edge were performed at 2 K with an $\pm$ 5 T external magnetic field applied along the beam axis to reach saturation. The Ir $L_{2,3}$-edge XAS and XMCD were collected at beamline P09 of PETRA III at DESY \cite{Strempfer_JSR_2013}, where the powder was mixed and pressed into a low-Z BN material to produce a pellet for transmission measurements. The Ir $L_{2,3}$ XMCD measurements were performed at 5 K by fast-switching the beam helicity between left and right circular polarization \cite{Strempfer_AIP_2016}. To align the magnetic domains and correct for nonmagnetic artifacts, an external magnetic field of $\pm$ 5 T was applied parallel and antiparallel to the incident beam wave vector $k$ using a 6T/2T/2T vector magnet.

The magnetization and electrical transport measurements were performed on a Quantum Design's Physical Property Measurement System (PPMS) coupled with a VSM-head. The magnetization as a function of magnetic field [M(H)] curves were carried after zero-field-cool (ZFC) the sample, while the magnetization as a function of temperature [M(T)] curves were measured in both ZFC and field-cooled (FC) modes. The electrical transport data were carried out using platinum wires and silver paste for the contacts to produce a four-contact configuration. 

\section{Results}

\begin{figure}
\begin{center}
\includegraphics[width=0.8 \textwidth]{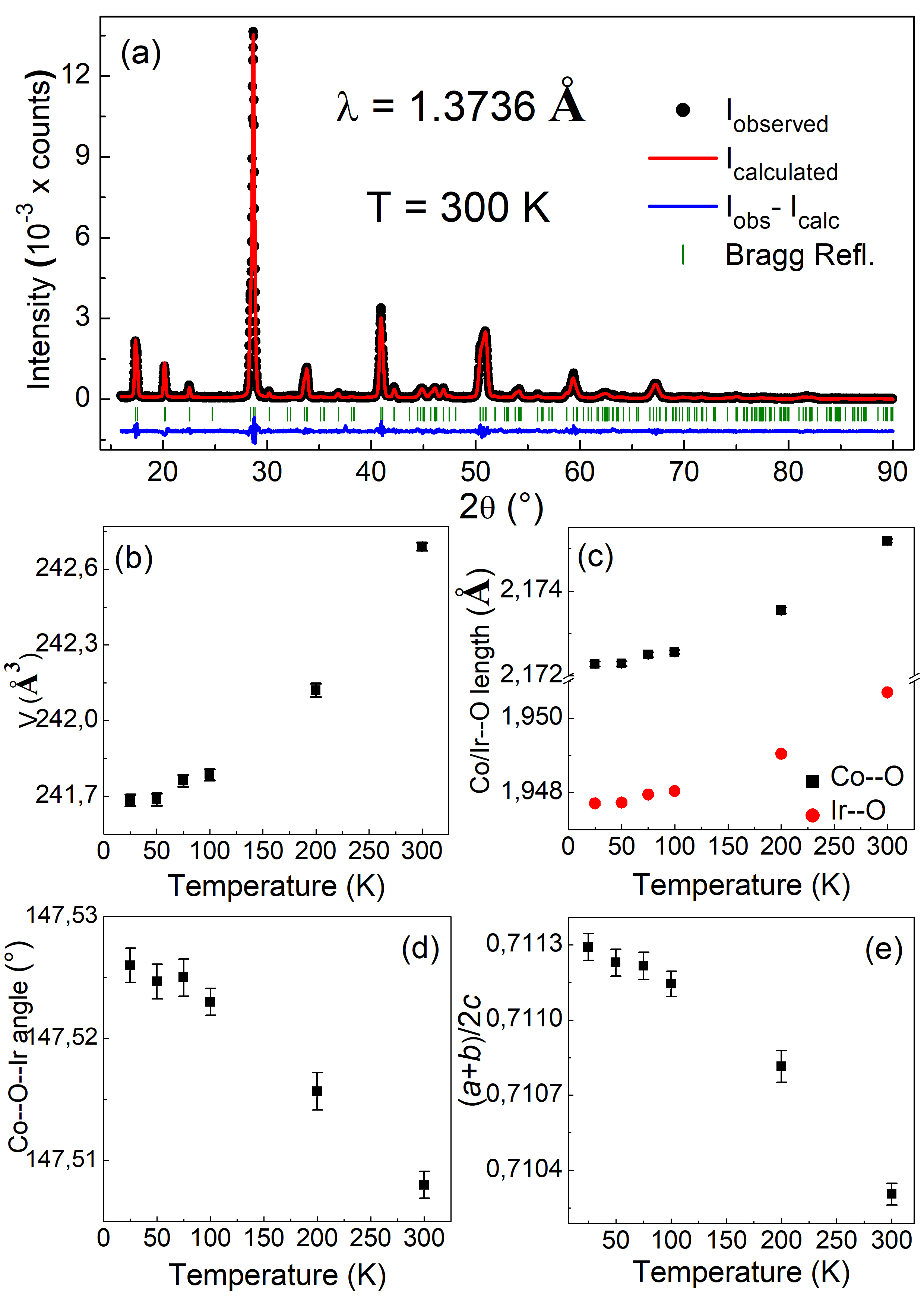}
\end{center}
\caption{(a) Rietveld refinement fitting of the room temperature SXRD of LCCIO. The vertical lines represent the Bragg reflections for the $P2_{1}/n$ space group. (b) Unit cell volume (V) as a function of $T$. (c) Average Co--O and Ir--O bond lengths as a function of $T$. (d) Average Co--O--Ir bond angle as a function of $T$. (e) ($a$+$b$)/2$c$ as a function of $T$.}
\label{Fig_XRD}
\end{figure}

Fig. \ref{Fig_XRD}(a) shows the SXRD pattern taken at 300 K, together with its Rietveld refinement in the $P2_{1}/n$ space group. The match between the observed and calculated patterns confirms that LCCIO grows in the monoclinic space group, in agreement with previous reports \cite{PRB2020}. The antisite disorder (ASD) at the Co/Ir sites is estimated from the refinement to be $\sim$6.7\%, which is within the values found for other 3$d$/Ir-based DPs \cite{Feng,Narayanan,PhysicaB,Haskel}. The SXRD data carried out at lower temperatures ($T$) show no evidence of any structural transition. The main parameters obtained from the Rietveld refinements are displayed in Table \ref{T1}. As Table \ref{T1} and Fig. \ref{Fig_XRD}(b) show, the unit cell volume ($V$) decreases with $T$, as expected. Although Figs. \ref{Fig_XRD}(b) and (c) seem to evidence some changes in the slope of curves for $V$ and Co/Ir--O bond lengths at the magnetic transition temperature, more data points would be necessary to ensure a conclusive interpretation. Besides the magnetic ordering, such changes in the slope of the curves may stem from the progressive freezing of phonons that occurs at low temperatures, causing the observed flattening of the structural evolution.

Interestingly, the average Co--O--Ir bond angle monotonically increases as $T$ decreases to 100 K, below which it it continues to increase, albeit at a lower rate. This suggests a tendency toward a constant value caused by the Co--Ir magnetic coupling (mediated by the intervening oxygen ions) that naturally strengthens at temperatures below the magnetic ordering. Such an increase of the Co--O--Ir angle may indicate a flattening of the oxygen octahedra, related to the fact that the reduction in lattice parameter $c$ is proportionally more pronounced than the changes in $a$ and $b$. This can be observed in Fig. \ref{Fig_XRD}(e) showing ($a$+$b$)/2$c$ as a function of $T$. This term, which is equal to 1 for a cubic system, can be used as a measure of the tetragonal distortion. As it can be seen, this parameter gets closer to 1 as $T$ decreases, corresponding to a decrease in the tetragonal distortion. 

\begin{table*}
\centering
\caption{Main results obtained from the Rietveld refinements of the SXRD data. The number in parenthesis represents the standard uncertainties in the last decimal digits.}
\label{T1}
\begin{tabular}{l|llllll}
\hline \hline
$T$ (K) & 300 & 200 & 100 & 75 & 50 & 25 \\

\hline

$a$ (\AA) & 5.52494(18) & 5.51829(25) & 5.51343(21) & 5.51264(21) & 5.51200(22) & 5.51168(22) \\

$b$ (\AA) & 5.60612(21) & 5.60675(30) & 5.60824(25) & 5.60903(26) & 5.60860(26) & 5.60919(27) \\

$c$ (\AA) & 7.83541(43) & 7.82556(64) & 7.81955(51) & 7.81877(55) & 7.81788(54) & 7.81738(54) \\

$V$ (\AA$^{3}$) & 242.69(2) & 242.12(3) & 241.79(2) & 241.76(2) & 241.69(2) & 241.68(2) \\

$\beta$ ($^{\circ}$) & 90.033(2) & 90.027(2) & 90.032(2) & 90.032(2) & 90.027(2) & 90.028(2) \\

$\langle$Co-O$\rangle$ (\AA) & 2.17518(5) & 2.17354(7) & 2.17256(5) & 2.17250(6) & 2.17229(6) & 2.17228(6) \\

$\langle$Ir-O$\rangle$ (\AA) & 1.95069(3) & 1.94904(5) & 1.94804(5) & 1.94794(5) & 1.94773(5) & 1.94771(5) \\

$\langle$Co-O-Ir$\rangle$ ($^{\circ}$) & 147.508(1) & 147.516(2) & 147.523(1) & 147.525(2) & 147.525(1) & 147.526(1) \\

\hline 

$R_{wp}$ (\%) & 13.6 & 12.9 & 12.9 & 13.4 & 13.4 & 13.4 \\

$R_{p}$ (\%) & 8.7 & 8.2 & 8.7 & 9.0 & 9.1 & 9.2 \\

\hline \hline
\end{tabular}
\end{table*}

\begin{figure}
\begin{center}
\includegraphics[width=0.75 \textwidth]{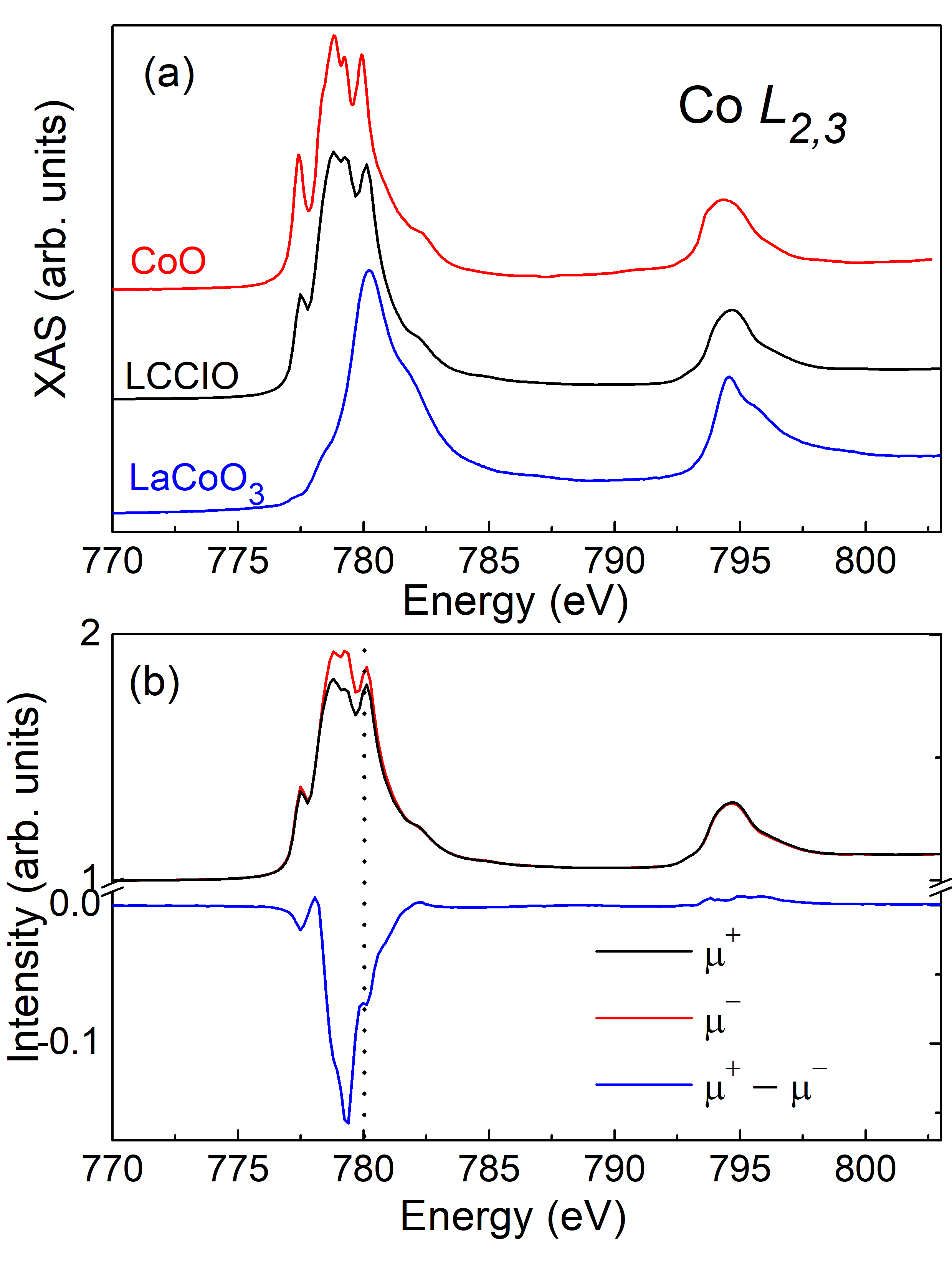}
\end{center}
\caption{(a) Co $L_{2,3}$-edge XAS spectra of LCCIO, CoO (Co$^{2+}$) and LaCoO$_3$ (Co$^{3+}$). (b) XMCD of LCCIO, taken at 2 K with $H$ = 5 T. The photon spin was aligned parallel ($\mu^{+}$, black) and antiparallel ($\mu^{-}$, red) to $H$, and the difference spectra are shown in blue. The dotted line is a guide to the eye.}
\label{Fig_CoL}
\end{figure}

Fig. \ref{Fig_CoL}(a) shows the Co $L_{2,3}$-edges XAS of LCCIO together with those of CoO and LaCoO$_3$, measured as references for Co$^{2+}$ and Co$^{3+}$ oxidation states, respectively \cite{Cho,Raveau}. The spectral features of LCCIO are closer to that of CoO and to that of other Co$^{2+}$-based DPs \cite{Cho,Burnus}. For the XMCD signal depicted in Fig. \ref{Fig_CoL}(b), the black and red curves stand respectively for parallel ($\mu^{+}$) and antiparallel ($\mu^{-}$) alignments between the photon spin and the external magnetic field ($H$), and the difference spectra $\Delta\mu = \mu^{+} - \mu^{-}$ is represented by the blue line. The shape of the XMCD curve is also very similar to that of Co-based DPs for which the bivalent Co state dominates \cite{Cho,Burnus,PRB2019}. Particularly, the $L_3$-edge XMCD exhibits a shoulder at $\sim$780 eV that is characteristic of Co$^{2+}$ \cite{Cho,Burnus} but is usually absent in Co$^{3+}$-based systems \cite{Agrestini}. These results strongly indicate a nearly Co$^{2+}$ state in LCCIO, although the presence of small amounts of Co$^{3+}$ can not be totally ruled out.

As Fig. \ref{Fig_CoL}(b) shows, the Co $L_2$ XMCD signal is weakly positive while the Co $L_3$ is largely negative, giving evidence of a significant orbital contribution to Co magnetic moment. In order to get a quantitative estimate of the spin and orbital magnetic moments, we performed sum rules calculations developed by B. T. Thole \textit{et al.} \cite{Thole} and P. Carra \textit{et al.} \cite{Carra}, which provide the following orbital and spin contributions to the magnetization.
\begin{equation}
m_{l}=- \frac{4\int_{L_3+L_2}(\mu^+-\mu^-)d\omega}{3\int_{L_3+L_2}(\mu^++\mu^-)d\omega}N_{h}, \label{Eq1}
\end{equation} 
\begin{equation}
m_{s} = -\frac{6\int_{L_3}(\mu^+-\mu^-)d\omega-4\int_{L_3+L_2}(\mu^+-\mu^-)d\omega}{\int_{L_3+L_2}(\mu_++\mu_-)d\omega}\times N_{h}\left(1+ \frac{7\langle T_z\rangle}{2\langle S_z \rangle}\right)^{-1},  \label{Eq2}\\
\end{equation}
where $m_{l}$ and $m_{s}$ are the angular and spin magnetic moments in units of $\mu_B$/atom, $S_z$ denotes the projection along $z$ of the spin magnetic moment, $N_h$ represent the number of empty 3$d$ states, $T_z$ denotes the magnetic dipole moment, and $L_2$ and $L_3$ represent the integration ranges. 

Here, we assumed $T_z$  to be negligible compared to $S_z$, as usual for 3$d$ TM ions in octahedral symmetry \cite{Teramura,Groot}. Thus, using $N_h$ = 3, which is an approximated atomistic value for Co$^{2+}$ state corresponding to 3 holes in the 3$d$ level, we obtain $m_{l}$ $\simeq$ 0.14 $\mu_B$ and $m_{s}$ $\simeq$ 0.29 $\mu_B$, revealing a significant orbital moment, as expected. Considering sources of deviations such as electronic interactions, imprecision in the integral calculations, and the $N_h$ value assumed, we estimate an uncertainty of 10$\%$ on these values \cite{Teramura,Groot} (see table \ref{T2}).

\begin{table}
\caption{Co and Ir BR, orbital ($m_l$), spin ($m_s$) and total ($m_t$) magnetic moments obtained from the Co and Ir $L_{3,2}$ XAS and XMCD.}
\label{T2}
\begin{tabular}{c|ccccc}
\hline \hline

Ion & $m_l$ ($\mu_B$) & $m_s$ ($\mu_B$) & $m_t$ ($\mu_B$) & $m_l/m_s$ & BR \\

\hline

Co & 0.141$\pm$0.014 & 0.286$\pm$0.029 & 0.427$\pm$0.032 & 0.49 & 5.5 \\

Ir & -0.022$\pm$0.002 & -0.015$\pm$0.002 & -0.037$\pm$0.003 & 1.47 & 4.6 \\

\hline \hline

\end{tabular}
\end{table}

The orbital and spin magnetic moments obtained are relatively close but smaller than those reported by Min-Cheol Lee \textit{et al.} for La$_2$CoIrO$_6$ ($m_l$ = 0.18 $\mu_B$, $m_s$ = 0.31 $\mu_B$ \cite{Cho}. Such difference could be related to some internal field-induced contribution since for La$_2$CoIrO$_6$; the neighboring Ir ions are mainly in the tetravalent state. In contrast, for LCCIO they are majoritarian in pentavalent $J$ = 0 states, as will be discussed later. In any case, the total magnetic moment obtained for LCCIO from the sum rules, $m_t$ $\simeq$ 0.43 $\mu_B$, falls significantly below the $\sim$3 $\mu_B$ value expected for Co$^{2+}$. This disparity indicates an AFM arrangement of the Co ions, in agreement with other analogue CoIr-based DPs where the small Co contribution to the magnetisation is attributed tospin canting \cite{Narayanan,Lee}.

\begin{figure}
\begin{center}
\includegraphics[width=0.75 \textwidth]{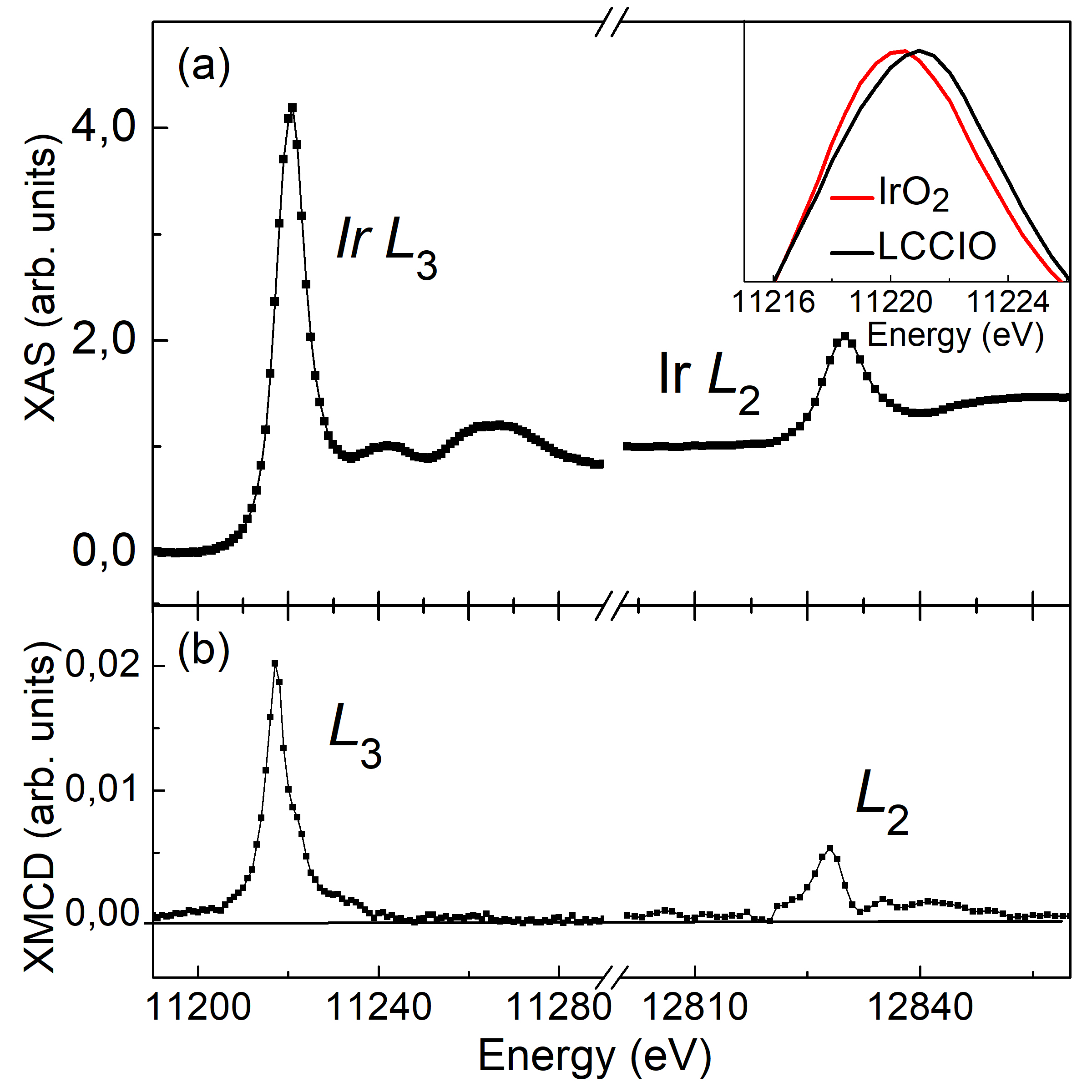}
\end{center}
\caption{(a) Ir $L_{3}$-edge XAS of LCCIO and and IrO$_2$ standard. Inset shows a magnified view of the absorption edges to compare their positions in energy. (b) $L_{2,3}$-edge XMCD spectra of LCCIO, measured at 5 K and under $H$ = 5 T.}
\label{Fig_IrL}
\end{figure}

Fig. \ref{Fig_IrL}(a) shows the Ir $L_{2,3}$-edge XAS of LCCIO. Its spectral lineshape is qualitatively very similar to that of both Ir$^{4+}$- and Ir$^{5+}$-based DPs \cite{PRB2020,Agrestini,Kolchinskaya,PRB2023,Fabbris}. Indeed, the similarity of the $L_3$-edge XAS is characteristic of Ir and other 5$d$ ions, being related to the diffuse 5$d$ valence orbitals \cite{PRB2020,Liu}. However, a closer inspection of the $L_3$ absorption threshold for LCCIO, as compared with that of IrO$_2$ measured as a reference for Ir$^{4+}$ state [inset of Fig. \ref{Fig_IrL}(a)], reveals a shift of $\sim$0.8 eV between the white lines. Although such displacement lies within the instrumental resolution ($\sim$1.5 eV), it indicates a tendency toward Ir$^{5+}$ in our system, but some small amount of Ir$^{4+}$ may be also present, since in octahedral coordination it is usually observed a shift of $\sim$1 eV or more toward higher energies for the Ir$^{5+}$ $L_{3}$ XAS spectra with respect to that of Ir$^{4+}$  \cite{PRB2020,Haskel,Agrestini}. This, in turn, suggests the presence of a small fraction of Co$^{3+}$ and/or oxygen vacancy.  

We can further scrutinize the tendency toward Ir$^{5+}$ in LCCIO and, additionally, get further insight into the strength of the SOC by computing the branching ratio BR = $I_{L_3}$/$I_{L_2}$, where $I_{L_3}$ and $I_{L_2}$ are the integrated white line intensities calculated from the $L_3$ and $L_2$ absorption edges, respectively. The value here found, BR $\simeq$ 4,6, is larger than that observed for Ir$^{4+}$ but similar to those of Ir$^{5+}$ in DPs \cite{Haskel}. This tends to confirm that Ir ions are majoritarian in the pentavalent state. Importantly, this value is much larger than the statistical BR $\simeq$ 2 observed for Ir in metals \cite{Croft} and alloys \cite{Kohori}, suggesting a strong SOC. From the BR we can estimate the ground-state expectation value of the angular part of the SOC by using the equation BR = (2 + $\langle$ L $\cdot$ S $\rangle$/$N_h$)/(1 - $\langle$ L $\cdot$ S $\rangle$ / $N_h$). Assuming $N_h$ $\simeq$ 6 (\textit{i.e.} an Ir formal valence close to +5), we have $\langle$ L $\cdot$  S $\rangle$ $\simeq$ 2.8 $\hbar^{2}$. This high value tends to corroborate with a $j_{eff}$ description for the Ir electronic configuration in LCCIO \cite{Haskel,Veenendaal,Clancy}.

Fig. \ref{Fig_IrL}(b) shows the Ir $L_{2,3}$-edge XMCD signal for LCCIO, measured at 5 K with $H$ = 5 T. Differently from previous report on the same compound \cite{PRB2020}, a non-negligible $L_3$ signal is observed. Such discrepancy is most likely related to the fact that in Ref. \cite{PRB2020} the XMCD was carried out at a different condition, with $H$ = 0.9 T at 60 K. Our $M(T)$ and $M(H)$ data, along with element-specific XMCD hysteresis curves reported for the resemblant compounds La$_2$CoIrO$_6$ \cite{Kolchinskaya}, La$_{1.5}$Ca$_{0.5}$CoIrO$_6$ and La$_{1.5}$Ba$_{0.5}$CoIrO$_6$ \cite{PRB2023}, suggest that the Ir moments in LCCIO are far from reaching saturation with $H$ = 0.9 T at 60 K. Interestingly, it can be noticed a relatively large Ir $L_2$ signal. For other Ir$^{5+}$-based DPs, usually the $L_3$ and $L_2$ signals have opposite signs, leading to a vanishing XMCD integrated intensity and, consequently, to a small orbital moment \cite{Agrestini,Kolchinskaya}. In contrast, for LCCIO, the $L_3$ and $L_2$ XMCD signals have the same sign, resulting in a noticeable orbital moment. It is important to mention that the sign of the Ir XMCD signal is opposite to that of Co, further indicating an AFM coupling between these ions.

We performed the sum rules calculations for a quantitative estimate of the Ir orbital moment. Assuming $N_h$ = 6 in Eq. \ref{Eq1} yields $m_l$ = -0.022 $\mu_B$. Using $\langle$$T_z$$\rangle$/$\langle$$S_z$$\rangle$ = 0.056 as previously obtained from configuration interaction calculations for Ir$^{5+}$-based DPs \cite{Irifune,Haskel,Veenendaal} we get $m_s$ = -0.015 $\mu_B$ (using $\langle$$T_z$$\rangle$ = 0 would lead to $m_s$ = -0.018 $\mu_B$, a difference of 20\%), resulting in a large orbital to spin ratio typical of 5$d$-based materials presenting strong SOC \cite{Yi,Fujiyama}. Although the magnetic moments here obtained are much smaller than those found for Ir$^{4+}$ \cite{PRB2023,Kolchinskaya}, they are somewhat larger than those usually reported for Ir$^{5+}$ \cite{Irifune,Haskel,Veenendaal}. The recent discussion concerning excitonic magnetism for Ir$^{5+}$ in some DPs [4, 5, 6, 15] indicates that non-negligible magnetism in Ir$^{5+}$ can not be completely ruled out in our sample. However, we believe that any interpretation of our data in this way would be merely speculative since the Ir $L_{2,3}$-edge XAS suggests the presence of some tetravalent Ir ions. The relatively large moment found is most likely related to the presence of magnetic Ir$^{4+}$. Again, the magnetic moments obtained from sum rules are just rough estimates. Considering sources of imprecision such as the calculated XAS and XMCD integrated intensities, the $N_h$ and $\langle$$T_z$$\rangle$ values assumed, we estimate an uncertainty of $\sim$ 20\% on these values. 

\begin{figure}
\begin{center}
\includegraphics[width=0.75 \textwidth]{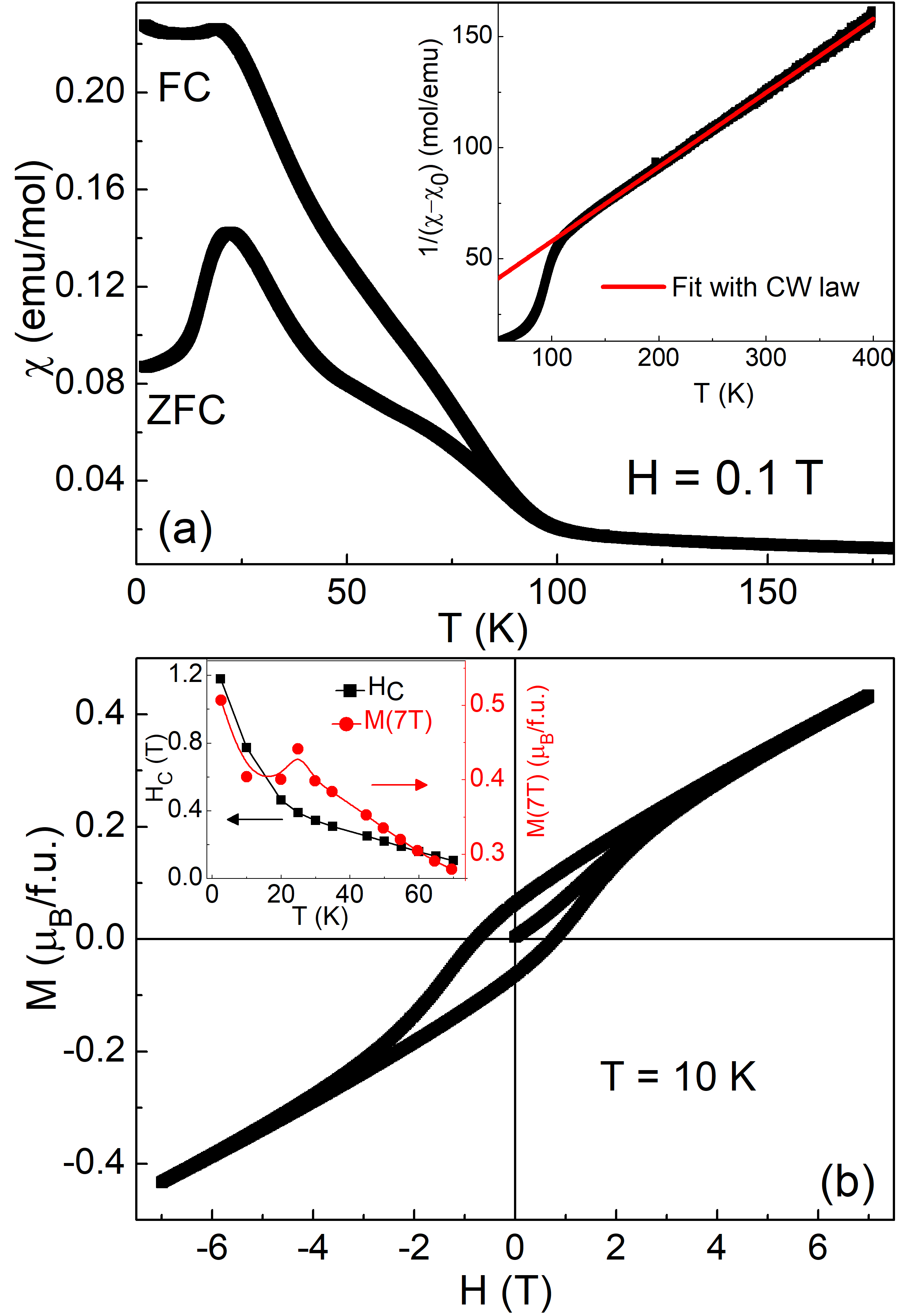}
\end{center}
\caption{(a) ZFC-FC $\chi_{dc}$ M(T) curves for LCCIO measured with $H$ = 0.1 T. The inset shows 1/($\chi$-$\chi_0$) and the fit of its PM region with the CW law. (b) M(H) curve measured at 10 K. The inset shows the evolution of $H_C$ and $M$(7T) with temperature. The lines are guides for the eye.}
\label{Fig_PPMS}
\end{figure}

The ZFC and FC M(T) curves of LCCIO, measured with $H$ = 0.1 T, are displayed in Fig. \ref{Fig_PPMS}(a). The FC curve exhibits an FM-like ordering at $\sim$87 K followed by a rough plateau-like behavior below $\sim$24 K. In contrast, the ZFC curve shows an anomaly at $\sim$87 K corresponding to the material's conventional magnetic ordering and a pronounced cusp at $\sim$24 K. Previous AC susceptibility measurements performed on this compound confirmed that the low-temperature cusp is related to the emergence of a spin glass (SG) phase, making of this a re-entrant SG system \cite{PRB2020}. 

The inset shows 1/($\chi$-$\chi_0$) $vs$ temperature, where the $\chi_0$ term represents a $T$-independent contribution due to the combined effects of the core diamagnetic susceptibility and a Van Vleck paramagnetic susceptibility \cite{Cao,Cao2,Agrestini}. The straight line represents the best fit of the paramagnetic region of the curve with the Curie-Weiss (CW) law, from which we obtain $\chi_0$ = 3.9 $\times$ 10$^{-4}$ emu/mol$\cdot$Oe, with this value being in the range usually found for Ir$^{5+}$-based DPs \cite{Cao2,Agrestini,Jansen}. The negative sign of the CW temperature extracted from the fit, $\theta_{CW}$ = -74 K, indicates a predominance of AFM coupling in LCCIO, in agreement with other DPs containing a 3$d$ TM ion and Ir$^{5+}$ such as LaSrNiIrO$_6$, Sr$_2$FeIrO$_6$, and Sr$_2$CoIrO$_6$, for which it is attributed an AFM arrangement of the 3$d$ ions, with the Ir$^{5+}$ being presumably in the $J$ = 0 state \cite{Hayward,Page,Agrestini}. Despite our low-temperature magnetization being somewhat larger than what is observed in the materials mentioned above, it remains considerably lower than the value expected for FM or even FIM arrangements. The FM-like qualitative features observed in our study are most likely related to spin canting in an AFM superstructure at the Co site, reminiscent of findings reported for LaSrCoIrO$_6$ \cite{Narayanan,Kolchinskaya}.

The fit of the PM region with the CW law yields an effective magnetic moment $\mu_{eff}$ = 4.9 $\mu_B$/f.u, which is somewhat larger than the value expected from a spin-only approximation for Co$^{2+}$, $\mu$ = 3.9 $\mu_B$ \cite{Ashcroft}, and considering a $J$ = 0 state for the Ir$^{5+}$. Notwithstanding its 3$d$ character, Co$^{2+}$ usually exhibits unquenched orbital contribution to its magnetic moment in octahedral coordination \cite{Raveau}. Furthermore, our XAS data suggest small fractions of Ir$^{4+}$ and Co$^{3+}$, helping to explain the $\mu_{eff}$ value found. In its turn, the presence of mixed valences and ASD at Co/Ir sites would lead to competing magnetic phases and, consequently, to the emergence of the SG phase observed at lower temperatures \cite{PRB2020}.

Fig. \ref{Fig_PPMS}(b) shows the M(H) curve measured at 10 K. As can be seen, the loop shows some hysteresis characteristic of FM or FIM. But again, the small magnetization values, together with the linear dependence of the curve with $H$ at high fields, suggest AFM coupling, with the FM-like behavior possibly coming from spin canting at the Co moment \cite{Narayanan,Lee}. The inset of Fig. \ref{Fig_PPMS}(b) shows the evolution of the coercive field ($H_C$) with temperature, as well as the magnetization at $H$ = 7 T, $M$(7T). A large coercivity at low temperatures can be noticed, which can be anticipated as due to strong magnetic anisotropy caused by significant orbital contribution from both Co and Ir ions \cite{Feng,Agrestini,PRB2023,Escanhoela}. The changes in the slopes of  $M$(7T) and $H_C$ curves below the freezing temperature of the SG state ($\sim24$ K) indicate intrinsic blocking mechanisms related to local frustration effects of the glassy magnetism \cite{Campbell}. 

\begin{figure}
\begin{center}
\includegraphics[width=0.75 \textwidth]{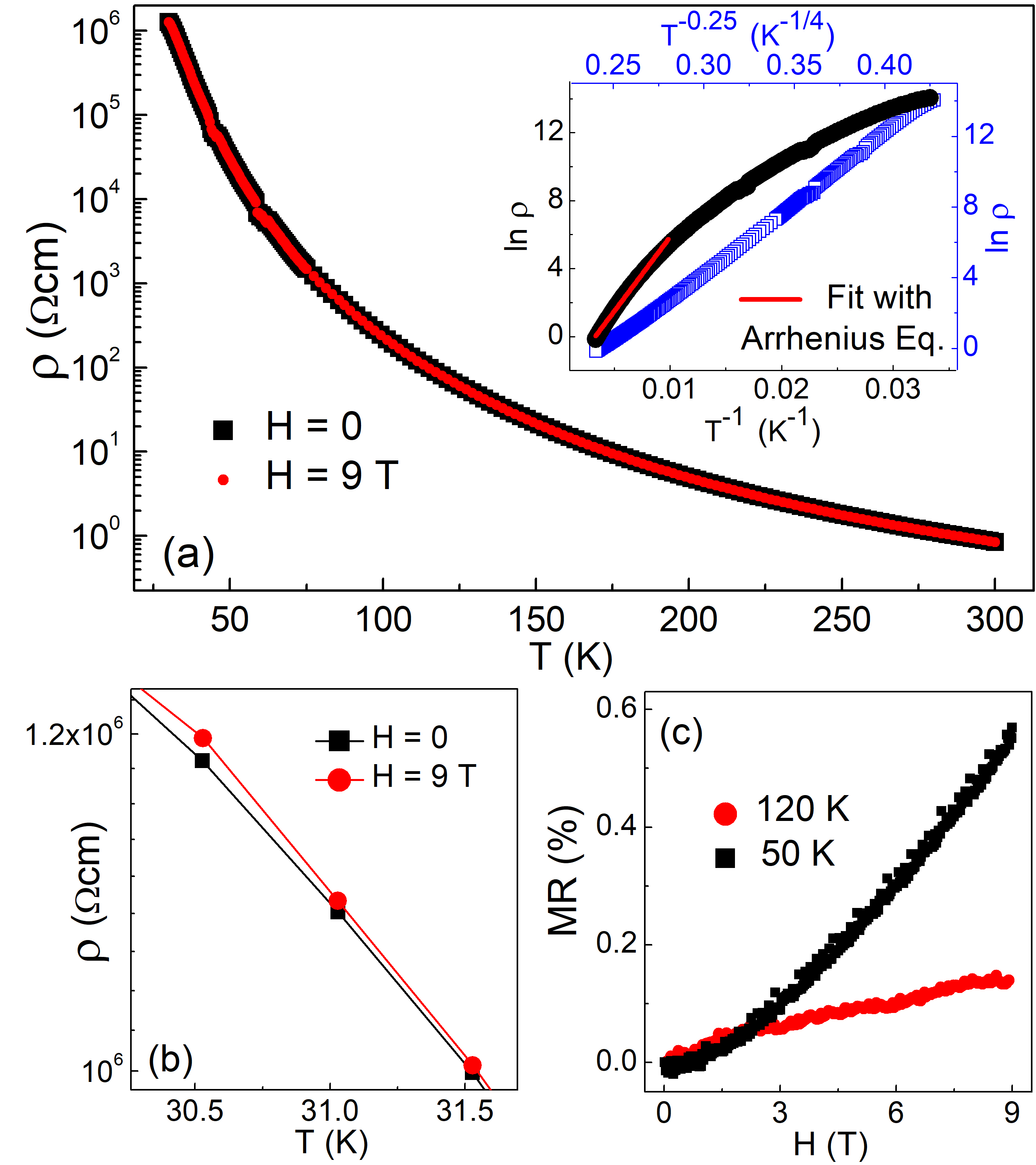}
\end{center}
\caption{(a) Electrical resistivity ($\rho$) as a function of temperature, measured with $H$ = 0 and 9 T. (b) Magnified view of the low $T$ region, highlighting the positive MR. (c) MR as a function of $H$, measured at 50 K and 120 K.}
\label{Fig_rho}
\end{figure}

The electrical resistivity measurements show semiconductor-like behavior for LCCIO, as observed in Fig. \ref{Fig_rho}(a). At room temperature, the resistivity is $\rho$ $\simeq$ 0.8 $\Omega\cdot$cm and increases exponentially with decreasing temperature down to $\sim$30 K, below which it exceeds the instrumental limit of detection. The inset of Fig. \ref{Fig_rho}(a) shows $ln\rho$ vs. $T^{-1}$, from which the high temperature data ($T$ $>$ 150 K) was fitted with the Arrhenius equation $\rho(T)$ $\propto$ exp($E_{g}/2k_{B}T$), where $E_g$ is the energy bandgap and $k_B$ is the Boltzmann constant. The best-fit yields $E_g$ $\simeq$ 0.13 eV, a value similar to that of other Ir-based DPs \cite{Narayanan,Jansen,Esser}. However, the fitting of $ln\rho$ on a $T^{-1/4}$ scale, also shown at the inset, results in a significantly better agreement.  This indicates that the electrical transport of LCCIO is better described by a variable-range hopping mechanism \cite{Mott} rather than by a thermally activated model.

The application of a magnetic field leads to a subtle increase of $\rho$, which is unnoticed in the scale of Fig. \ref{Fig_rho}(a) but can be visualized in the zoom-in view of the low-temperature region, Fig. \ref{Fig_rho}(b). This positive magnetoresistance occurs along the whole investigated temperature interval but is larger below $T_N$. This can be observed in Fig. \ref{Fig_rho}(c), where isothermal $\rho$ vs $H$ curves measured at 50 K ($<$ $T_N$) and 120 K ($>$ $T_N$) are depicted. At 50 K, the magnetoresistance, $MR(H)$ = 100$\times$[$\rho(H)$ - $\rho(0)$]/$\rho(0)$, is of $\sim$0.6\% for $H$ = 9 T, while at 120 K it is about five times smaller.

\section{Discussion}

A recent study of first-principles calculations on La$_2$CoIrO$_6$ indicates that the energy of Ir $t_{2g}$ states reside in between those of Co $t_{2g}$ and $e_g$ states. In contrast, the Ir $e_g$ states lie farther above, making these last orbitals irrelevant to our discussion \cite{Ganguly}. From the experimental side, we came to a similar conclusion using XAS, XMCD and magnetometry experiments on La$_{1.5}A_{0.5}$CoIrO$_6$ ($A$ = Ba, Ca) DPs, for which the Co $e_g$ -- Ir $t_{2g}$ ($j_{eff}$ = 1/2) AFM coupling seems to be the most relevant path of hybridization between Co$^{2+/3+}$ and Ir$^{4+}$ \cite{PRB2023}. Also recently, a study of XAS and magnetotransport on strained Sr$_{2}$CoIrO$_{6}$ films has shown that for compressively strained films the Co$^{3+}$ moments points perpendicular to the surface of the film within an AFM superstructure. Moreover, applying a magnetic field along this direction results in a positive magnetoresistance that gradually increases with increasing the field strength. This enhancement is attributed to the progressive rotation of the moments, originally oriented in an antiparallel direction, towards the alignment with the applied magnetic field \cite{Esser}. 

\begin{figure}
\begin{center}
\includegraphics[width=0.60 \textwidth]{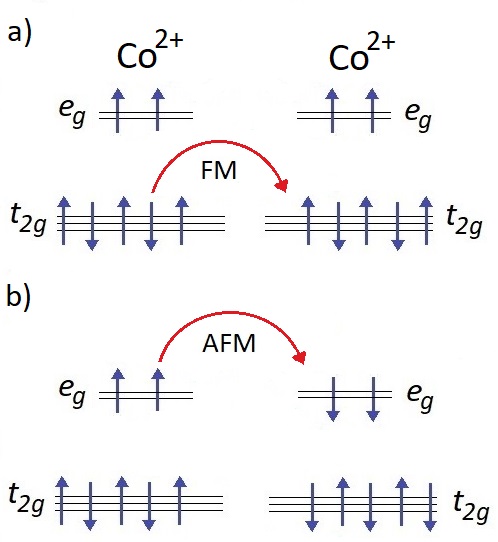}
\end{center}
\caption{Schematic diagram of the mechanisms of the virtual electron hopping between Co ions via (a) the $t_{2g}$ orbitals, resulting in FM coupling, and (b) the $e_{g}$ orbitals, leading to AFM. The intermediate Ir and O ions are omitted for simplicity.}
\label{Fig_Draw}
\end{figure}

Although the Co$^{2+}$/Ir$^{5+}$ valence states in LCCIO differ from those found in  La$_2$CoIrO$_6$ (Co$^{2+}$/Ir$^{4+}$), La$_{1.5}$(Ba,Ca)$_{0.5}$CoIrO$_6$ (Co$^{2+/3+}$/Ir$^{4+}$) and Sr$_2$CoIrO$_6$ (Co$^{3+}$/Ir$^{5+}$), we can use the information discussed above to address the mechanisms of hybridization in our compound. Co$^{2+}$ (3$d^{7}$) has two completely filled and one half filled $t_{2g}$ orbital, while both its $e_g$ orbitals are half filled. For Ir, the SOC lifts the $t_{2g}$ degeneracy into a fourfold $j_{eff}$ = 3/2 and a twofold $j_{eff}$ = 1/2 state. In LCCIO, the Ir valence is close to +5 (5$d^{4}$), with the $j_{eff}$ = 3/2 orbitals full and the $j_{eff}$ = 1/2 ones empty. This means that the non-magnetic Ir (here neglecting any possible excitonic magnetism for Ir$^{5+}$) is just an intermediate path for the exchange interactions between the Co ions. Thereby, the virtual electron hopping through the pseudolinear Co--O--Ir--O--Co or the orthogonal Co--O--O--Co paths can occur via the Co $t_{2g}$ or $e_g$ orbitals, as schematically depicted in Fig. \ref{Fig_Draw}. Importantly, despite the fact that the $t_{2g}$-$t_{2g}$ hybridization is, in principle, precluded by symmetry in a cubic perovskite, it becomes possible with the tilting and rotation of the oxygen octahedra \cite{Serrate,Sami}, as is the case for the distorted monoclinic structure of LCCIO. 

The upper panel of Fig. \ref{Fig_Draw} shows the exchange interaction between Co $t_{2g}$ orbitals, while the bottom panel depicts the exchange coupling via the Co $e_g$ orbitals. As can be seen, the $t_{2g}$--$t_{2g}$ hybridization should result in FM coupling between the Co ions, whereas for the Co $e_g$--$e_g$ interaction, an AFM coupling is predicted. The small magnetization values observed by magnetometry and XMCD suggest that FM is unlikely in LCCIO, which favors the AFM structure for the Co ions. This is corroborated by the positive magnetoresistance found in the electrical transport data, since for an FM arrangement of the Co ions through $t_{2g}$ orbitals, one should expect an increase in the conductivity when $H$ is applied. Conversely, in the presence of AFM coupling, a strong external field acts to align the spins, thus inhibiting the virtual electron hopping through $e_g$ orbitals and consequently increasing the material's resistivity. All together, our results indicate that the Co $e_g$--$e_g$ AFM hybridization, mediated by the intervening O 2$p$ and Ir $t_{2g}$ ($j_{eff}$ = 1/2) orbitals, is the most relevant in Co$^{2+}$/Ir$^{5+}$ systems.

\section{Summary}

In summary, we thoroughly investigated the structural, electronic, and magnetic properties of polycrystalline LaCaCoIrO$_6$ (LCCIO) using SXRD, M(T) and M(H) measurements, Co and Ir $L_{2,3}$-edge XAS and XMCD, and electrical transport. LCCIO crystallizes in the monoclinic $P2_1/n$ space group with $\sim$6.7\% of ASD at the Co/Ir sites. The XAS and XMCD results indicate a nearly Co$^{2+}$/Ir$^{5+}$ configuration, with possibly small fractions of Co$^{3+}$ and Ir$^{4+}$ ions. The weak FM-like behavior observed in the M(T) and M(H) curves is most likely related to spin canting at the Co sites. The positive magnetoresistance observed in the electrical transport data suggests an AFM arrangement of the Co ions via hybridization between its $e_g$ orbitals and the Ir $t_{2g}$ ($j_{eff}$ = 1/2) orbitals.

\textbf{ACKNOWLEDGEMENTS } 

This work was supported by the Brazilian funding agencies: Funda\c{c}\~{a}o Carlos Chagas Filho de Amparo \`{a} Pesquisa do Estado do Rio de Janeiro (FAPERJ) [Nos. E-26/202.798/2019 and E-26/211.291/2021], Funda\c{c}\~{a}o de Amparo \`{a}  Pesquisa do Estado de Goi\'{a}s (FAPEG) and Conselho Nacional de Desenvlovimento Cient\'{\i}fico e Tecnol\'{o}gico (CNPq) [No. 400633/2016-7]. We thank Diamond Light Source for time on beamline I06 under proposal MM29620-1. We acknowledge DESY (Hamburg, Germany), a member of the Helmholtz Association HGF, for providing experimental facilities. Parts of this research were carried out at beamline P09 under proposal I-20200348, and we would like to thank J. Bergtholdt and O. Leupold for assistance in setting up the experiment and 6T/2T/2T vector magnet, respectively.

\end{document}